\documentclass[twocolumn,showpacs,preprintnumbers,amsmath,amssymb,superscriptaddress]{revtex4}

\usepackage{graphicx}
\usepackage{dcolumn}
\usepackage{bm}

\begin{document}

\preprint{APS/126-QED}

\title{Yb$_{2}$Pt$_{2}$Pb: Magnetic frustration in the Shastry-Sutherland lattice}

\author{M. S. Kim}
\affiliation{Condensed Matter Physics and Materials Science
Department, Brookhaven National Laboratory, Upton, New York
11973-5000}\affiliation{Department of Physics, University of
Michigan, Ann Arbor, Michigan 48109-1120}
\author{M. C. Bennett}
\affiliation{Department of Physics, University of Michigan, Ann
Arbor, Michigan 48109-1120}\affiliation{Department of Physics and
Astronomy, Stony Brook University, Stony Brook, New York 11794-3800}
\author{M. C. Aronson}
\affiliation{Condensed Matter Physics and Materials Science
Department, Brookhaven National Laboratory, Upton, New York
11973-5000} \affiliation{Department of Physics, University of
Michigan, Ann Arbor, Michigan 48109-1120}\affiliation{Department
of Physics and Astronomy, Stony Brook University, Stony Brook,
New York 11794-3800}

\date{\today}

\begin{abstract}

We have synthesized single crystals of Yb$_{2}$Pt$_{2}$Pb, which
crystallize in the layered U$_{2}$Pt$_{2}$Sn-type structure, where
planes of Yb ions lie on a triangular network. We report here the
first results of magnetization, specific heat, and electrical
resistivity experiments. The lattice constants and high
temperature magnetic susceptibility indicate that the Yb ions are
trivalent, while Schottky peaks in the specific heat show that
the ground state is a well isolated doublet. Significant magnetic
anisotropy is observed, with the ratio of susceptibilities
perpendicular and parallel to the magnetic planes differing by as
much as a factor of 30 at the lowest temperatures.
Antiferromagnetic order occurs at a N\'eel temperature $T_{\rm
N}$=2.07 K, a transition temperature which is more than an order
of magnitude smaller than the mean field interactions reflected
by the in-plane Weiss temperature. Further evidence for short
ranged magnetic fluctuations is found in the magnetic
susceptibility and electrical resistivity, which have broad peaks
above $T_{\rm N}$, and in the slow development of the magnetic
entropy at $T_{\rm N}$. Our experiments indicate that
Yb$_{2}$Pt$_{2}$Pb is a quasi-two dimensional and localized
moment system, where strong magnetic frustration may arise from
the geometry of the underlying Shastry-Sutherland lattice.

\end{abstract}

\maketitle

\section{Introduction}
Recently, the series of ternary compounds with composition
R$_{2}$T$_{2}$M (R=rare earths or actinides; T=transition metals;
M=Cd, In, Sn, and Pb) has attracted much attention because of the
diversity of its collective phenomena, including ferromagnetic Kondo
lattice,\cite{gorden} Kondo semiconductor,\cite{plessis} valence
fluctuation,\cite{kaczorowski,giovannini,dogan,dhar} non-Fermi
liquid,\cite{dhar,bauer} and heavy fermion ground
states.\cite{dhar,hauser,havela}

Most of this series crystallize in the tetragonal
Mo$_{2}$FeB$_{2}$-type structure (space group $P4/mbm$), which is an
a ordered derivative of the U$_{3}$Si$_{2}$-type. A smaller number
of compounds crystallize in a distorted variant of this structure,
the U$_{2}$Pt$_{2}$Sn-type structure (space group $P4_{2}/mnm$),
which is a superstructure of the
U$_{3}$Si$_{2}$-type.\cite{hulliger,pereira,gravereau,pottgen0,pottgen1,pottgen2}
Both structures are intrinsically layered, with two types of layers
which are alternately stacked along the $c$ axis. The first layer
type contains only R atoms, while the second layer type contains
only T and M atoms. In the first layer type, the R atoms are
arranged in a triangular motif, giving each R atom a single nearest
neighbor. This structure is suggestive that planes of magnetic
dimers may be formed, a necessary ingredient for novel spin liquid
states, where frustration competes with order.

The crystal structure of the rare earth sublattices in both the
Mo$_{2}$FeB$_{2}$ and U$_{2}$Pt$_{2}$Sn-types suggest that they may
be rare examples of the two dimensional Shastry-Sutherland lattice,
which has received substantial theoretical attention due to its
exact dimerization and its spin liquid ground
state.\cite{shastry,miyahara} Currently, there are only a few known
materials known to be examples of Shastry-Sutherland systems. One
such system is SrCu$_{2}$(BO$_{3}$)$_{2}$, which has a layered
structure where each Cu$^{2+}$ ion is antiferromagnetically coupled
to a single nearest neighbor and to four next nearest neighbors,
with each unit cell containing two of these dimers, which are
mutually orthogonal. Features of this dimerized lattice include the
absence of magnetic order, the formation of a spin gap at low
temperature, and the appearance of plateaux in the magnetization
with 1/4, 1/8, and 1/10 of the full saturated
moment,\cite{kageyama1,kageyama2} as different ordered states emerge
from the initially frustrated spin liquid through the application of
magnetic fields.\cite{miyahara} Recently, similar plateaux have also
been found in the magnetization of single crystalline TmB$_{4}$ and
ErB$_{4}$ which form in the same space group as the layered
Mo$_{2}$FeB$_{2}$-type structure, with the rare earth atoms forming
the Shastry-Sutherland
network.\cite{etourneau,michimura,yoshii,iga}. Given this
relationship between the Shastry-Sutherland lattice and the
Mo$_{2}$FeB$_{2}$, U$_{2}$Pt$_{2}$Sn, and related crystal structure
types, it is important to identify additional examples with
different moment degeneracy and enhanced degrees of
two-dimensionality.

It is known that a number of different members of the
R$_{2}$T$_{2}$M series with R=Ce,Yb, or U form in either the
Mo$_{2}$FeB$_{2}$ or U$_{2}$Pt$_{2}$Sn structures. In order to
ascertain their suitability as exemplars of the Shastry-Sutherland
lattice, it is important to establish whether the fundamental
attributes of magnetic anisotropy, dimerization, and geometrical
frustration are present. So far, most studies on these latter
compounds have been conducted on polycrystalline samples, precluding
a definitive answer to this question. We have recently succeeded in
synthesizing single crystals of Yb$_{2}$Pt$_{2}$Pb, which
crystallize in the tetragonal U$_{2}$Pt$_{2}$Sn-type
structure.\cite{pottgen2} Here, we present a first report of the
anisotropic magnetization and magnetic susceptibility, the specific
heat, and the electrical resistivity. We will argue here that the Yb
moments in Yb$_{2}$Pt$_{2}$Pb lie on a lattice which is equivalent
to the Shastry-Sutherland lattice.

\section{Experimental Details}

\begin{table}[t]
\caption{\label{tab:table1}Results of the crystal structure
refinement for Yb$_{2}$Pt$_{2}$Pb.}
\begin{ruledtabular}
\begin{tabular}{ll}
Space group & $P4_{2}/mnm$(No. 136) \\
Lattice parameters & $a=7.7651(6)$ \AA \\
 & $c=7.0207(7)$ \AA \\
Formula units per cell & $Z=4$ \\
Cell volume & 423.327 \AA$^{3}$ \\
Calculated density & 14.80 g/cm$^{3}$ \\
\multicolumn{2}{l}{Conventional Rietveld reliability factors\footnotemark[1]:}\\
\multicolumn{2}{l}{$R_{\rm P}=0.0976$, $R_{\rm WP}=0.1290$, $R_{\rm exp}=0.1145$} \\
\multicolumn{2}{l}{$R_{\rm B}=0.0361$, $R_{\rm F}=0.0265$}  \\
Goodness of fit &  $\chi^{2}=1.28$ \\
\end{tabular}
\end{ruledtabular}
\footnotetext[1]{$R_{\rm P}$: Profile factor, $R_{\rm WP}$:
Weighted profile factor, $R_{\rm exp}$: Expected weighted profile
factor, $R_{\rm B}$: Bragg factor, $R_{\rm F}$: Crystallographic
factor}
\end{table}

\begin{table}[t]
\caption{\label{tab:table2}Crystallographic data for
Yb$_{2}$Pt$_{2}$Pb at room temperature.}
\begin{ruledtabular}
\begin{tabular}{cccccc}
&Wyckoff&&&\\
 Atom & Site & $x$ & $y$ & $z$
 & $U_{\rm eq}$\footnotemark[1](\AA$^{2}$)\\
\hline
Yb1 & $4f$ & 0.1771(8) & $x$ & 0 & 0.0074(23)\\
Yb2 & $4g$ & 0.3386(8) & $-x$ & 0 & 0.0055(21)\\
Pt & $8j$ & 0.3730(3) & $x$ & 0.2747(9) & 0.0066(9)\\
Pb & $4d$ & 0 & 1/2 & 1/4 & 0.0050(9)\\
\end{tabular}
\end{ruledtabular}
\footnotetext[1]{$U_{\rm eq}$ stands for the isotropic thermal
parameter defined as one-third of the trace of the orthogonalized
$U_{\rm ij}$ tensor.}
\end{table}

Single crystals of Yb$_{2}$Pt$_{2}$Pb were grown from a Pb flux,
forming in a rod-like morphology with a square cross-section.
Elements in the ration Yb:Pt:Pb = 5:4:40 were placed in an alumina
crucible and then sealed in an evacuated quartz tube. The
materials were initially held at 1180 $^{\circ}$C for four hours,
and subsequently cooled to 450 $^{\circ}$C at 7 $^{\circ}$C per
hour. Excess Pb was removed by centrifuging the tubes at 450
$^{\circ}$C. Laue photo images indicate that the crystallographic
$c$ axis ($\lbrack001\rbrack$) lies along the long axis of the
crystal, while the $a$ ($\lbrack100\rbrack$) and $b$
($\lbrack010\rbrack$) axes define the square cross-section of the
rod. Electron-probe microanalysis was carried out on polished
single crystals using a Cameca SX100 microprobe system with
elemental Yb, Pt, and Pb standards. The atomic ratio of Yb:Pt:Pb
is found to be $40.5\pm0.3\%:39.2\pm0.4\%:20.2\pm0.5\%$, and is
spatially uniform over the whole crystal surface. X-ray
diffraction was performed on a powder prepared from the single
crystals. The diffraction patterns were refined using Fullprof.
Refinement results and crystallographical data are listed in
Tables ~\ref{tab:table1} and \ref{tab:table2}, respectively. Both
confirm that Yb$_{2}$Pt$_{2}$Pb has the previously reported
U$_{2}$Pt$_{2}$Sn-type structure.\cite{pottgen2}

Magnetic susceptibility measurements were performed using a Quantum
Design Magnetic Property Measurement System (MPMS) and a Vibrating
Sample Magnetometer (VSM) in a Quantum Design Physical Property
Measurement System (PPMS) over temperature ranges of 1.8 K $<T<$ 300
K and 300 K $<T<$ 800 K, respectively. Magnetization measurements
were performed using the VSM in magnetic fields up to 14 T. The
electrical resistivity was measured by the conventional four-probe
method in the temperature range $0.4<T<300$ K. Specific heat
measurements were carried out for temperatures between 0.4 K and 100
K.

\section{Crystal structure}

\begin{figure}[t]
\begin{center}
\includegraphics[width=7.9cm]{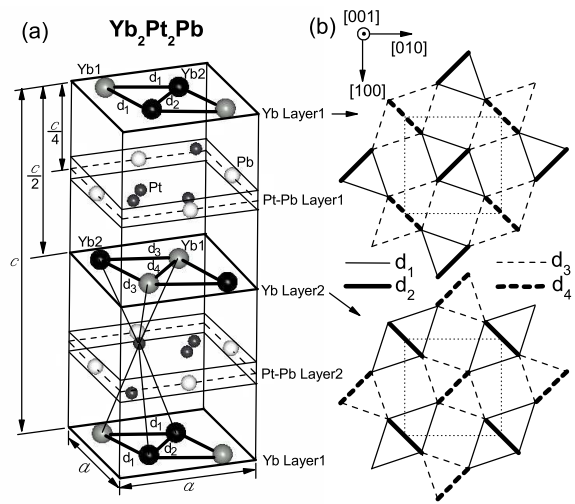}
\end{center}
\caption{\label{f1} (a) Schematic representation of the unit cell
of Yb$_{2}$Pt$_{2}$Pb crystallizing in the tetragonal
$P4_{2}/mnm$ (No. 136) structure. (b) Lattice structure of the Yb
ions in Yb layer 1 (top) and 2 (bottom). The Yb ions lie at the
vertices of the triangles and the dotted lines delineate the unit
cell.}
\end{figure}

In the tetragonal Mo$_{2}$FeB$_{2}$-type structure, T and M atoms of
R$_{2}$T$_{2}$M lie in the same plane, while T atoms in the
tetragonal U$_{2}$Pt$_{2}$Sn-type structure are displaced out of the
plane, as demonstrated in the crystal structure of
Yb$_{2}$Pt$_{2}$Pb shown in Fig.~\ref{f1}(a). Here, two of the four
Pt atoms on the middle of the Pt-Pb plane are shifted down (up) and
the other two close to the edge of the plane are shifted up (down)
in the Pt-Pb layer 1 (in the Pt-Pb layer 2) without any shift of Pb
atoms from the plane. These displacements result in two kinds of
Yb-Pt tetrahedra, with large and small heights as shown in
Fig.~\ref{f1}(a). The shifted Pt atoms force the Yb atoms to form
two different networks of isosceles triangles ($d_{1}$ = 3.9646
\AA~and $d_{2}$ = 3.5451 \AA~in Yb layer 1, and $d_{3}$ = 4.1960
\AA~and $d_{4}$ = 3.8890 \AA~in Yb layer 2), leading to a
superstructural distortion with a doubled lattice parameter $c$.
Among the rare earth and actinide intermetallics, only a few display
a similar superstructural
distortion.\cite{pottgen0,pottgen1,gravereau,pereira,hulliger,pottgen2}
On the other hand, in the Yb layer 1 and 2, Yb atoms are arranged in
the network of mixed rectangles and isosceles triangles as shown in
Fig.~\ref{f1}(b), implying that the two rare earth layers can
separately be considered topologically similar to Shastry-Sutherland
lattices.\cite{shastry} This suggests that magnetic frustration is
likely to be an important feature of the magnetic behavior in
Yb$_{2}$Pt$_{2}$Pb.

\begin{figure}[t]
\begin{center}
\includegraphics[width=7.5cm]{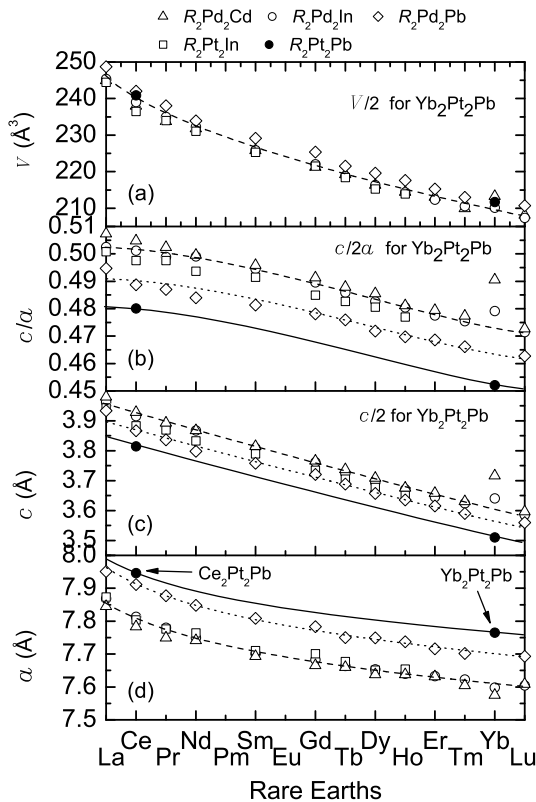}
\end{center}
\caption{\label{f2} Comparison of lattice parameters $a$ and $c$,
the ratio $c/a$, and the unit cell volume $V$ for R$_{2}$T$_{2}$M
(R=rare earths, T=Pd and Pt, M=Cd, In, and
Pb).\cite{giovannini,dogan,pottgen2,hulliger1,melnyk,pottgen3}
Solid, dotted, and dashed lines are guides for the eye.}
\end{figure}

Fig.~\ref{f2} compares the structural parameters for some members of
the R$_{2}$T$_{2}$M (R=rare earths; T=Pd and Pt; M=Cd, In, and Pb)
series~\cite{giovannini,dogan,pottgen2,hulliger1,melnyk,pottgen3}
which crystallize in the tetragonal Mo$_{2}$FeB$_{2}$-type
structure, except for Yb$_{2}$Pt$_{2}$Pb which has the
U$_{2}$Pt$_{2}$Sn-type structure. All parameters gradually decrease
as the R atoms are changed from La to Lu, due to the lanthanide
contraction. For comparison, $c/2$, $c/2a$, and $V/2$ are plotted
for Yb$_{2}$Pt$_{2}$Pb. For lattice parameters $a$ and $c$, we find
the relation of $a({\rm Pt, Pb})$$>$$a({\rm Pd, Pb})$$>$$a({\rm Pt,
In})$$\gtrsim$$a({\rm Pd, In})$$\gtrsim$$a({\rm Pd, Cd})$ and
$c({\rm Pd, Cd})$$\gtrsim$$c({\rm Pd, In})$$\gtrsim$$c({\rm Pt,
In})$$>$$c({\rm Pd, Pb})$$>$$c({\rm Pt, Pb})$. Considering atomic
radii $r$ ($r_{\rm Pt}$$\gtrsim$$r_{\rm Pd}$ and $r_{\rm
Pb}$$>$$r_{\rm In}$$\gtrsim$$r_{\rm Cd}$), this indicates that the
increase of the sum of the radii of T and M atoms underlies the
increase of the lattice parameter $a$ and the decrease of the
lattice parameter $c$. The inverse dependence of $c$ compensates the
dependence of $a$ resulting in a much weaker variation in the unit
cell volume with rare earth atom, as shown in Fig.~\ref{f2}(a). We
note that the anomalously large values of $c$ and $c/a$ found in
Yb$_{2}$Pd$_{2}$M (M=Cd and In) are caused by the intermediate
valence of Yb in these compounds.\cite{dogan,dhar,giovannini,bauer}
In contrast, the values of $c$ and $c/a$ in Yb$_{2}$Pt$_{2}$Pb are
consistent with those of other compounds with trivalent rare earth
ions. This indicates that Yb in Yb$_{2}$Pt$_{2}$Pb is in a stable
trivalent Yb state, without mixed valence character. From the
unusually low value of $c/a$ found in Yb$_{2}$Pt$_{2}$Pb, we
anticipate strongly anisotropic behavior.

Among the R$_{2}$Pt$_{2}$Pb (R=rare earth) series, only
Ce$_{2}$Pt$_{2}$Pb~\cite{pottgen3} and Yb$_{2}$Pt$_{2}$Pb have
previously been reported. Note that Ce$_{2}$Pt$_{2}$Pb crystallizes
in the tetragonal Mo$_{2}$FeB$_{2}$-type structure, while
Yb$_{2}$Pt$_{2}$Pb forms in the distorted U$_{2}$Pt$_{2}$Sn-type
variant. One might thus expect that R$_{2}$Pt$_{2}$Pb compounds
comprised of light rare earths R may crystallize in the tetragonal
Mo$_{2}$FeB$_{2}$-type structure, while the heavy rare earths
crystallize in the tetragonal U$_{2}$Pt$_{2}$Sn-type structure. A
similar effect was observed in the R$_{2}$Au$_{2}$In
series,\cite{hulliger} where only the heavy rare earths R=Tm and Lu
crystallize in the tetragonal U$_{2}$Pt$_{2}$Sn-type structure.
Considering the lanthanide contraction, this indicates that the
small size of the heavy rare earth atoms plays an important role in
enabling the superstructural distortion. Specifically, we suggest
that the displacement of the T atoms in R$_{2}$T$_{2}$M with the
tetragonal U$_{2}$Pt$_{2}$Sn-type structure only occurs for small R
atoms. This proposal is supported by the observations that the
compounds with divalent Yb (Yb$_{2}$Cu$_{2}$In and
Yb$_{2}$Au$_{2}$In)~\cite{tsujii,giovannini2} and the compounds with
mixed valent Yb (Yb$_{2}$Pd$_{2}$Cd, Yb$_{2}$Pd$_{2}$In, and
Yb$_{2}$Pd$_{2}$Sn),\cite{dogan,dhar,giovannini,bauer} where the Yb
atoms all have relatively large size, instead crystallize in the
Mo$_{2}$FeB$_{2}$-type structure and not in the
U$_{2}$Pt$_{2}$Sn-type structure. Conversely, it seems that Yb atoms
in the tetragonal Mo$_{2}$FeB$_{2}$-type structure prefer divalent
or mixed valent states to the trivalent state, while Yb atoms in the
tetragonal U$_{2}$Pt$_{2}$Sn-type structure prefer trivalent states.
We note that Yb$_{2}$Pt$_{2}$Pb is the only one of the reported Yb
compounds of R$_{2}$T$_{2}$M series which is trivalent. Recently,
composition dependent magnetic order has been found in compounds of
the series Yb$_{2}$Pd$_{2}$In$_{1-x}$Sn$_{x}$ for $x=0.6$ and 0.8,
intermediate between the mixed valent endpoints Yb$_{2}$Pd$_{2}$In
and Yb$_{2}$Pd$_{2}$Sn.\cite{bauer} Also, applying pressures between
1 and 4 GPa causes Yb$_{2}$Pd$_{2}$Sn to order
magnetically.\cite{muramatsu} These observations of magnetic order
imply that the Yb ions are trivalent. It would follow, then, that
the crystal structures of these compounds may well be distorted from
the tetragonal Mo$_{2}$FeB$_{2}$-type structure, possibly into the
U$_{2}$Pt$_{2}$Sn-type structure found in Yb$_{2}$Pt$_{2}$Pb. It
would be interesting to test for such a structural change in these
magnetically ordering compounds.

\section{Physical properties}

\subsection{Experimental results}

\begin{figure}[t]
\begin{center}
\includegraphics[width=7.4cm]{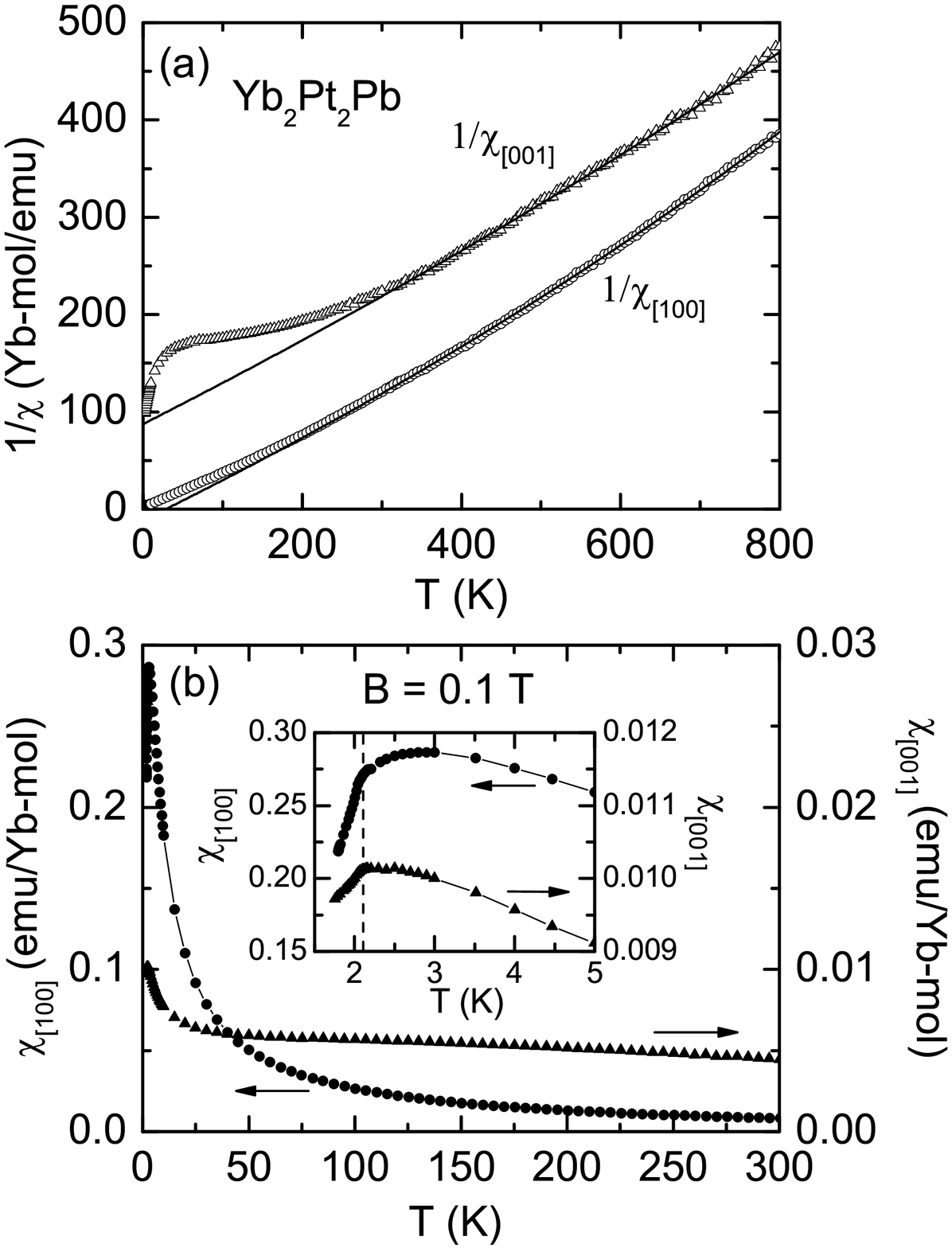}
\end{center}
\caption{\label{f3} (a) Temperature dependences of the inverse
susceptibilities $1/\chi_{\lbrack100\rbrack}$ ($\circ$) and
$1/\chi_{\lbrack001\rbrack}$ ($\vartriangle$) up to 800 K. The
magnetic susceptibility at 300 K $<T<$ 800 K was measured in 2 T.
The solid lines represent a fit with a modified Curie-Weiss law at
300 K $<T<$ 800 K (see text)(b) Temperature dependence of magnetic
susceptibilities $\chi_{\lbrack100\rbrack}$ ($\bullet$) and
$\chi_{\lbrack001\rbrack}$ ($\blacktriangle$) for
$B$$\parallel$$\lbrack100\rbrack$ and
$B$$\parallel$$\lbrack001\rbrack$, respectively, below 300 K in 0.1
T. The inset shows an enlarged plot of $\chi_{\lbrack100\rbrack}$
and $\chi_{\lbrack001\rbrack}$ with $T<5$ K.}
\end{figure}

Fig.~\ref{f3}(a) shows the inverse of the magnetic
susceptibilities 1/$\chi_{\lbrack100\rbrack}$ and
1/$\chi_{\lbrack001\rbrack}$, measured with a magnetic field
$B=0.1$ T along $\lbrack100\rbrack$ and $\lbrack001\rbrack$,
respectively, below 800 K. The data between 300 K and 800 K are
well described by a modified Curie-Weiss law
($\chi=\chi_{0}+C/(T-\theta)$), which gives $\chi_{0}=-0.0006$
emu/Yb-mol, a Weiss temperature $\theta_{\lbrack100\rbrack}=28$ K,
and an effective moment $\mu_{\rm eff}=4.42$ $\mu_{\rm B}$ for
$1/\chi_{\lbrack100\rbrack}$, while $\chi_{0}=-0.0004$
emu/Yb-mol, $\theta_{\lbrack001\rbrack}=-217$ K, and $\mu_{\rm
eff}=4.54$ $\mu_{\rm B}$ for $1/\chi_{\lbrack001\rbrack}$. The
effective moments deduced from $1/\chi_{\lbrack100\rbrack}$ and
$1/\chi_{\lbrack001\rbrack}$ are very close to 4.54 $\mu_{\rm B}$,
as expected for free Yb$^{3+}$ ions. This indicates that the Yb
moments in Yb$_{2}$Pt$_{2}$Pb are well localized and trivalent at
high temperatures, consistent with the analysis of the crystal
structure.

Fig.~\ref{f3}(a) shows that both 1/$\chi_{\lbrack100\rbrack}$ and
1/$\chi_{\lbrack001\rbrack}$ increasingly deviate from Curie-Weiss
behavior below ~300 K, and that this deviation is particularly
marked for 1/$\chi_{\lbrack001\rbrack}$. As indicated in
Fig.~\ref{f3}(b), $\chi_{\lbrack001\rbrack}$ is nearly independent
of temperature down to 30 K, with magnitude of only
$\sim$$5\times10^{-3}$ emu/Yb-mol, displaying a residual tail at the
lowest temperatures. For $1/\chi_{\lbrack100\rbrack}$, the deviation
from the Curie-Weiss law is less pronounced, and is only detected
below ~150 K. The inset of Fig.~\ref{f3}(b) shows
$\chi_{\lbrack100\rbrack}$ and $\chi_{\lbrack001\rbrack}$ below 5 K.
A broad maximum is found for  $\chi_{\lbrack001\rbrack}$ and
especially  $\chi_{\lbrack100\rbrack}$ between 2-3 K, while a sharp
cusp-like anomaly is found in $\chi_{\lbrack001\rbrack}$ and a
weaker anomaly in $\chi_{\lbrack100\rbrack}$, indicating the onset
of magnetic order at 2.07 K.

\begin{figure}[t]
\begin{center}
\includegraphics[width=7.5cm]{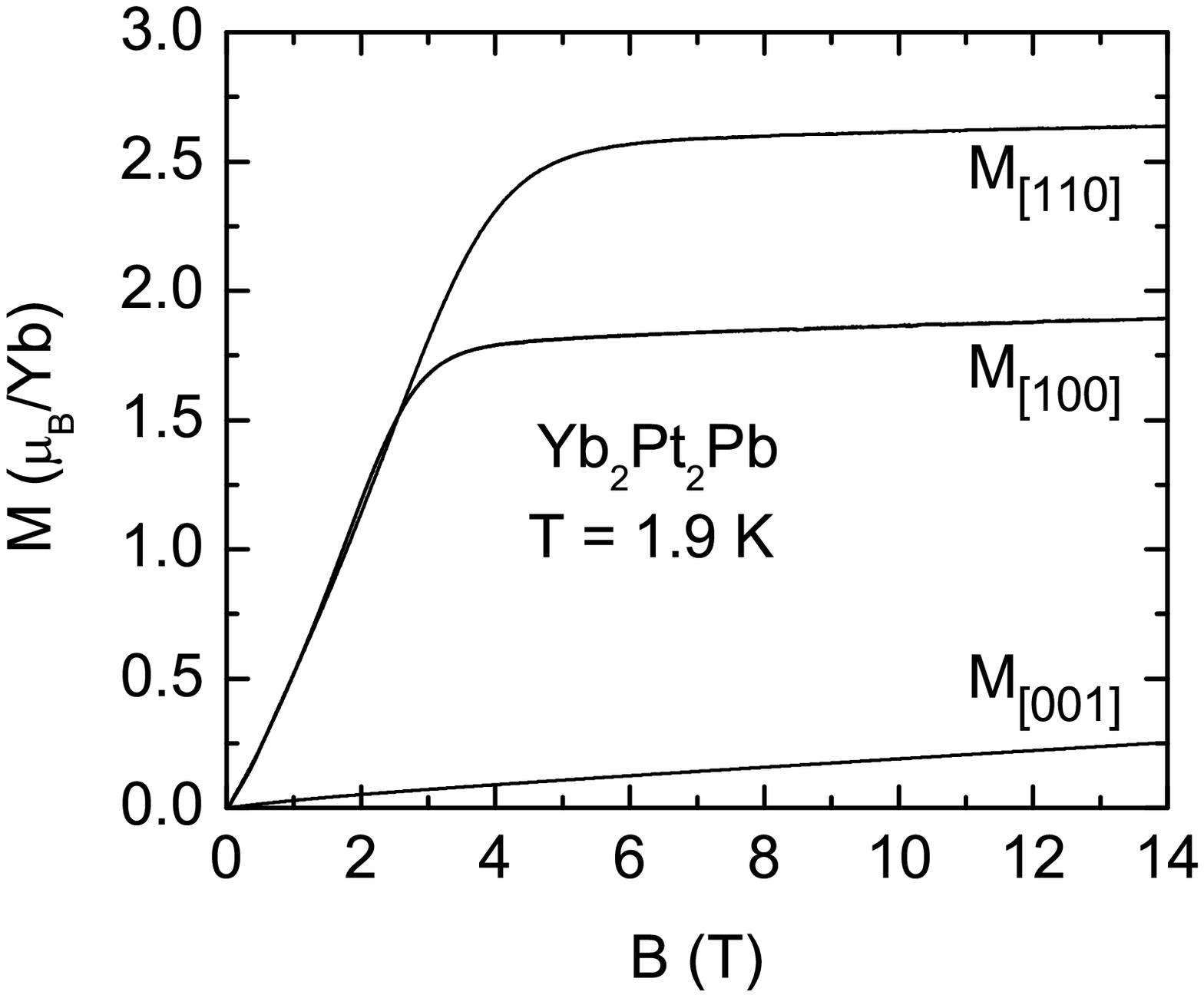}
\end{center}
\caption{\label{f4} (a) Magnetic field dependences of the
magnetizations $M_{\lbrack001\rbrack}$, $M_{\lbrack100\rbrack}$, and
$M_{\lbrack110\rbrack}$ for $B$$\parallel$$\lbrack001\rbrack$,
$B$$\parallel$$\lbrack100\rbrack$, and
$B$$\parallel$$\lbrack110\rbrack$, respectively.}
\end{figure}

The magnetic susceptibility is strongly anisotropic, with
$\chi_{\lbrack100\rbrack}/\chi_{\lbrack001\rbrack}\sim30$ at low
temperatures, a value which is even larger than that found in
R$_{2}$Cu$_{2}$In (R=Gd-Tm).\cite{fischer} Fig.~\ref{f4}(a) shows
the field dependence of the magnetizations $M_{\lbrack110\rbrack}$,
$M_{\lbrack100\rbrack}$, and $M_{\lbrack001\rbrack}$, measured at
1.9 K with the magnetic field oriented along the different principal
directions, $\lbrack110\rbrack$, $\lbrack100\rbrack$, and
$\lbrack001\rbrack$. It is clear that the magnetic hard axis is
along $\lbrack001\rbrack$, since here $M_{\lbrack001\rbrack}$
reaches only 0.25 $\mu_{\rm B}$/Yb in fields as large as 14 T. In
contrast, $M_{\lbrack110\rbrack}$ and $M_{\lbrack100\rbrack}$
gradually increase and then saturate at 2.6 $\mu_{\rm B}$ and 1.9
$\mu_{\rm B}$ above 5 T and 3 T, respectively. This last observation
indicates that the $\lbrack110\rbrack$ is the easy direction in the
plane, while the limited anisotropy between the $\lbrack110\rbrack$
and $\lbrack100\rbrack$ directions shows that this anisotropy is
very small compared to the anisotropy which confines the moments to
the layers themselves.

\begin{figure}[t]
\begin{center}
\includegraphics[width=7.5cm]{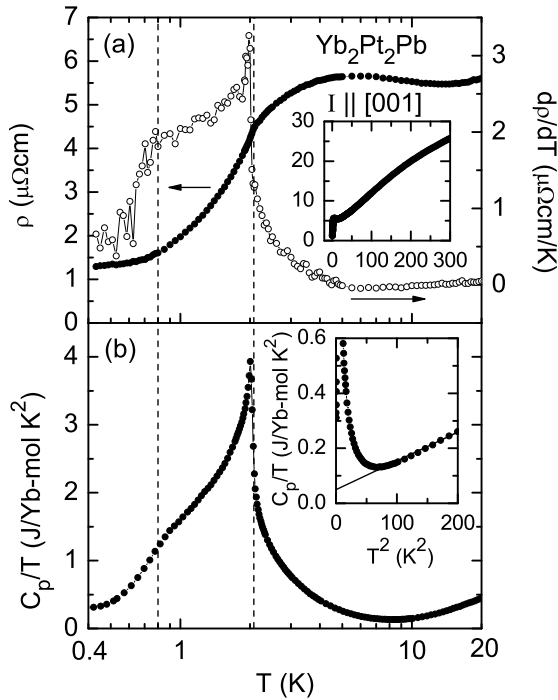}
\end{center}
\caption{\label{f5} (a) Logarithmic temperature dependence of the
resistivity $\rho$ ($\bullet$) and the temperature derivative of
the resistivity $d\rho/dT$ ($\circ$) below 20 K. The inset shows
the temperature dependence of $\rho$ over an expanded temperature
range. (b) Logarithmic temperature dependence of the specific
heat divided by temperature $C_{\rm p}/T$ below 20 K. The inset
shows the plot of $C_{\rm p}/T$ vs $T$ at high temperatures.
Solid line in the inset indicates the fit of $C_{\rm
p}/T=\gamma_{h}+\beta T^{2}$ above 8 K.}
\end{figure}

The inset of Fig.~\ref{f5}(a) shows the temperature dependence of
the electrical resistivity $\rho$ with the current flowing along
$\lbrack001\rbrack$. The resistivity is definitively metallic,
gradually decreasing with a weak positive curvature from its initial
value of 26 $\mu\Omega$cm at 300 K. The resistivity of most
polycrystalline compounds from the R$_{2}$T$_{2}$M series is very
high, typically several hundreds of $\mu\Omega$cm at 300 K and
several tens of $\mu\Omega$cm even at the lowest temperatures. For
Yb$_{2}$Pt$_{2}$Pb, the much lower resistivity at 300 K indicates
very good crystal quality and a near optimal growth of our single
crystals. Fig.~\ref{f5}(a) shows that with decreasing temperature, a
shallow minimum is found in $\rho$ at 13 K, followed by a broad
maximum preceding the sudden drop at 2.07 K, the same temperature at
which magnetic order is detected in $\chi_{\lbrack100\rbrack}$ and
$\chi_{\lbrack001\rbrack}$. Additional evidence for magnetic order
is found in the specific heat divided by temperature $C_{\rm p}/T$,
where a sharp peak is found at 2 K, as shown in Fig.~\ref{f5}(b).
Remarkably, Fig.~\ref{f5}(b) demonstrates that the temperature
dependences of $C_{\rm p}/T$ and $d\rho/dT$ are virtually identical,
both above and below the 2.07 K ordering temperature. Both show
distinct shoulder-like anomalies near 0.8 K, terminating at the
lowest temperatures in a residual resistivity $\rho_{0}$ = 1.3
$\mu\Omega$cm and an electronic part of the specific heat $\gamma$ =
311 mJ/Yb-molK$^{2}$, respectively. Given the high quality of our
crystals, we believe that the shoulder-like feature is intrinsic to
Yb$_{2}$Pt${_2}$Pb and does not result from the inclusion of a
secondary phase. However, it proves impossible to fit these data
over a convincing range of temperatures using any of the
conventional expressions for spin waves, either with or without
gaps.

\begin{figure}[t]
\begin{center}
\includegraphics[width=7.5cm]{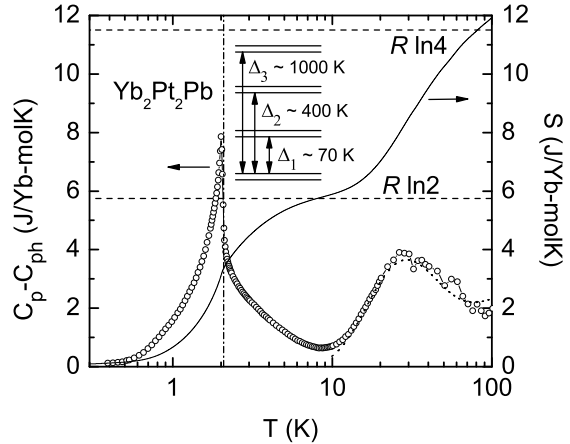}
\end{center}
\caption{\label{f6} The temperature dependence of the measured
specific heat $C_{\rm p}$ from which we have subtracted an estimated
phonon contribution $C_{\rm ph}$, and the associated entropy $S$.
The dotted line represents a fit to a Schottky expression. The
schematic diagram of the CEF levels from the fitting is presented in
the inset.}
\end{figure}

Fig.~\ref{f6} shows the magnetic contribution of the specific heat
$C_{\rm p}-C_{\rm ph}$, which is obtained by subtracting $C_{\rm
ph}$, the phonon contribution of the specific heat estimated from
the Debye model from the measured specific heat, $C_{\rm p}$. The
Debye temperature, $\theta_{\rm D}=184$ K is obtained by fitting
$C_{\rm p}/T=\gamma_{h}+\beta T^{2}$ above 8 K, as shown in the
inset of Fig.~\ref{f5}(b). The entropy $S$ is only 0.58$R\ln2$ at
2.07 K. With increasing temperature, $C_{\rm p}-C_{\rm ph}$ shows
a long tail up to 8 K at which a minimum is found and where the
entropy $S$ recovers its full value of $R\ln2$. The long tail
between 2.07 K and 8 K is consistent with the broad maxima in
$\chi_{\lbrack100\rbrack}$ and $\rho$ along $\lbrack001\rbrack$ in
indicating the presence of short ranged magnetic fluctuations in
this temperature range. The entropy $S$ remains approximately
constant between 10 and 15 K, indicating that magnetic order
develops from a well-separated doublet ground state. A broad
maximum is found near 30 K in $C_{\rm p}-C_{\rm ph}$, which can be
identified as a Schottky anomaly with an energy splitting scheme
of $\Delta_{1}\sim70$ K, $\Delta_{2}\sim400$ K, and
$\Delta_{3}\sim1000$ K between the four doublets expected for the
crystalline electric field (CEF) induced by the local $m2m$
symmetry of Yb sites in Yb$_{2}$Pt$_{2}$Pb, as shown in the inset
of Fig.~\ref{f6}. With increasing temperature, $S$ gradually
increases and eventually reaches $R\ln4$ around 70 K, as
expected, given the observation of the Schottky anomaly at 30 K.
This scale of crystal field splittings is comparable to that
found in inelastic neutron scattering measurements on
Yb$_{2}$Pd$_{2}$In.\cite{bauer}

\subsection{Discussion}

Our analysis of crystal structure trends in isostructural rare
earth compounds provides indirect evidence that the Yb moments in
Yb$_{2}$Pt$_{2}$Pb are trivalent. Magnetic susceptibility
measurements confirm this result, finding a conventional
paramagnetic state above $\sim$ 300 K which involves fluctuations
of the full Yb$^{3+}$ moment. Specific heat measurements
confirmed that the crystal electric field scheme is consistent
with the local site symmetry for the Yb$^{3+}$ moments. Further,
these measurements indicate that the ground state of
Yb$_{2}$Pt$_{2}$Pb involves a well isolated doublet for the Yb
moments.

Cusp-like anomalies in the magnetic susceptibilities
$\chi_{\lbrack100\rbrack}$ and especially
$\chi_{\lbrack001\rbrack}$ indicate the onset of
antiferromagnetic order at 2.07 K. Given the layered crystal
structure and the strong magnetic anisotropy which largely
confine the Yb moments to these layers, we might have expected
that intermoment interactions would be strong and
antiferromagnetic in the plane, and weaker between moments in
neighboring planes. In fact, the opposite situation is found. The
Weiss temperature $\theta_{\lbrack001\rbrack}$(= -217 K), found
when the field is perpendicular to the layers, is approximately
an order of magnitude larger than the in-layer Weiss temperature
$\theta_{\lbrack100\rbrack}$(= 28 K). We conclude, then, that the
anisotropy conferred on the Yb moments by the crystal electric
field must be the dominant factor in establishing magnetic
anisotropy in Yb$_{2}$Pt$_{2}$Pb. In this way, the competition
between the strong antiferromagnetic interactions between planes
and the weaker ferromagnetic interaction within the planes
results in a complex magnetic structure and its accompanying
fluctuations. Consequently, a sharp lambda-like anomaly is found
in the specific heat at 2.07 K, a temperature which is more than
two orders of magnitude lower than the inter-plane interaction
scale projected by the Weiss temperature
$\theta_{\lbrack001\rbrack}$= -217 K, and one order of magnitude
weaker than the ferromagnetic in-plane interactions implied by the
Weiss temperature $\theta_{\lbrack100\rbrack}$= 28 K.  The
temperature dependences of the specific heat and the temperature
derivative of the electrical resistivity are nearly identical,
both above and below the magnetic transition. This implies that
the resistivity is dominated by forward scattering, as expected
in a system with an abundance of small wave-vector critical
fluctuations, such as a ferromagnet.\cite{alexander} We conclude
that the magnetic structure of Yb$_{2}$Pt$_{2}$Pb may be rather
complex, with mixed antiferromagnetic and ferromagnetic
character, perhaps reflecting the geometrically frustrated nature
of the Shastry-Sutherland lattice. One possibility for the
zero-field structure would be a long-wave ferromagnetic spiral,
as has been proposed theoretically.\cite{albrecht,chung}

\begin{figure}[t]
\begin{center}
\includegraphics[width=8cm]{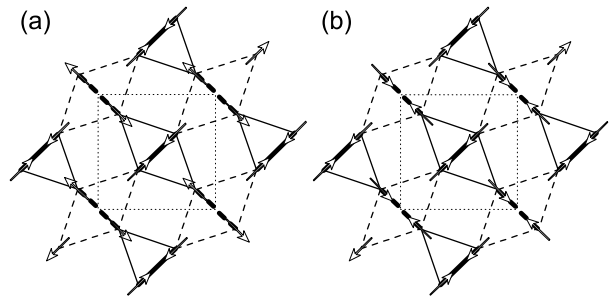}
\end{center}
\caption{\label{f7} Schematic representation of an expected
magnetic structure on Yb layers. (a) U$_{2}$Pt$_{2}$Sn type and
(b) GdB$_{4}$ type.}
\end{figure}

However, the underlying symmetries of the crystal lattice constrain
the magnetic structure of Yb$_{2}$Pt$_{2}$Pb. Yb atoms occupy two
inequivalent sites with local $m2m$ symmetry, lying at the
intersections of the (001), (110), and (1$\bar{1}$0) mirror planes.
It is well known that magnetic moments must lie in mirror planes or
be perpendicular to them. The very low values of
$\chi_{\lbrack001\rbrack}$ and $M_{\lbrack001\rbrack}$ indicate that
the magnetic moments of the Yb$^{3+}$ ions in Yb layers lie in the
(001) plane.  By analogy to the magnetic structures of
U$_{2}$Pt$_{2}$Sn and GdB$_{4}$,\cite{prokes,blanco} the Yb moments
are probably aligned along $\lbrack110\rbrack$ and
$\lbrack1\bar{1}0\rbrack$, in agreement with the anisotropy between
the saturation moments of $M_{\lbrack110\rbrack}$ and
$M_{\lbrack100\rbrack}$. Two possible magnetic structures which are
consistent with these constraints are shown in Fig.~\ref{f7}.

The onset of antiferromagnetic order from a strongly frustrated
paramagnetic state, as we observed in Yb$_{2}$Pt$_{2}$Pb, is
consistent with theoretical expectations for the Shastry-Sutherland
model. The strong magnetic anisotropy confirms that the moments lie
in the rare earth layers. In both layer types, the Yb moments lie on
a lattice of isosceles triangles ($d_{2}<d_{1}$ and $d_{4}<d_{3}$).
This separation of nearest Yb neighbors and next nearest Yb
neighbors results in a separation of the antiferromagnetic exchange
interactions ($J_{1}<J_{2}$ and $J_{3}<J_{4}$), leading to the
formation of well-defined pairs of Yb nearest neighbors in the two
rare earth layers(see Fig.~\ref{f1}(b)). In this way, the
arrangement of the rare earth ions in both layer types in
Yb$_{2}$Pt$_{2}$Pb are separately equivalent topologically to the
lattice of the Shastry-Sutherland model. Accordingly, the magnetic
structure inferred from the magnetization measurements
(Fig.~\ref{f7}) consists of sheets of orthogonal Yb dimers.

There is considerable evidence for the influence of frustration on
the magnetic properties of Yb$_{2}$Pt$_{2}$Pb. A universal feature
of magnetically frustrated systems is that magnetic order occurs at
temperatures which are much smaller than the scale of the
interactions revealed by the Weiss temperatures. By this measure,
Yb$_{2}$Pt$_{2}$Pb is highly frustrated, since the Weiss scale,
implies that the mean field interactions are more than two orders of
magnitude stronger than the 2.07 K ordering transition. The broad
maximum in $\chi_{\lbrack100\rbrack}$ just above the magnetic
ordering temperature is also suggestive of the short range magnetic
disorder common to systems with strong magnetic frustration. We have
considered the possibility that the Kondo effect is responsible for
these departures from ideal local moment behavior. Since it is a
single ion phenomenon, manifestations of the Kondo effect should
display no significant anisotropy. However, we see that the
susceptibility maximum is most pronounced in
$\chi_{\lbrack100\rbrack}$, suggesting that its origin must instead
be related to the intrinsic magnetic anisotropy of
Yb$_{2}$Pt$_{2}$Pb. In agreement with our results, similar maxima
are found in the anisotropic susceptibilities of
SrCu$_{2}$(BO$_{3}$)$_{2}$, TmB$_{4}$ and ErB$_{4}$ where the
magnetic ions also lie on the Shastry-Sutherland network, and where
no Kondo effect is likely.\cite{fisk,michimura} We conclude that the
broad maximum in $\chi_{\lbrack100\rbrack}$ reflects excess magnetic
fluctuations in the moment bearing planes, originating with the
geometrically frustrated Shastry-Sutherland lattice. These
fluctuations are also responsible for the slow recovery of the
magnetic entropy above the ordering temperature, detected as long
tails in the specific heat and electrical resistivity which extend
to temperatures as large as $\sim$ 10 K, i.e. five times the
transition temperature itself.

\begin{figure}[t]
\begin{center}
\includegraphics[width=7.5cm]{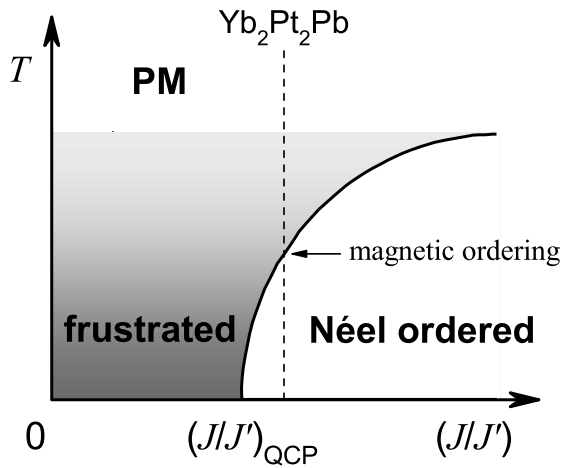}
\end{center}
\caption{\label{f8} Schematic phase diagram for the
Shastry-Sutherland lattice(see text).}
\end{figure}

The scenario in which long-range antiferromagnetic order emerges
from a frustrated magnetic liquid in Yb$_{2}$Pt$_{2}$Pb is
consistent with our expectations of the Shastry-Sutherland lattice.
Zero temperature calculations for spin-$\frac{1}{2}$ moments in the
two-dimensional Shastry-Sutherland lattice \cite{isacsson} predict a
quantum critical point when the ratio of the nearest neighbor $J'$
and next nearest neighbor $J$ exchange interactions ($J/J'$)$_{\rm
QCP}$$=$$0.7$. As shown in Fig.~\ref{f8}, the quantum critical point
separates a magnetically frustrated spin liquid for
$J/J'$$<$($J/J'$)$_{\rm QCP}$ from a N\'eel ordered ground state
with $J/J'$$>$($J/J'$)$_{\rm QCP}$. While there is little
theoretical guidance regarding the finite temperature properties of
the Shastry-Sutherland lattice, further increases of ($J/J'$) beyond
($J/J'$)$_{\rm QCP}$$=$$0.7$ are presumed to stabilize the N\'eel
ordered state at progressively higher temperatures, leading to the
phase line depicted qualitatively in Fig.~\ref{f8}. If the dominant
physics is that of the Shastry-Sutherland lattice, we assume that
Yb$_{2}$Pt$_{2}$Pb lies in this limit, where antiferromagnetic order
emerges from the spin liquid state. In order to establish this
conclusively, we must experimentally rule out other factors such as
structural modification, more distant neighbor exchange
interactions, and the interlayer exchange itself as possible drivers
of magnetic order in Yb$_{2}$Pt$_{2}$Pb.

\section{conclusion}

We have presented the magnetic, transport, and thermal properties of
single crystals of Yb$_{2}$Pt$_{2}$Pb crystallizing in the
tetragonal U$_{2}$Pt$_{2}$Sn-type structure, a superstructure
derived from the tetragonal Mo$_{2}$FeB$_{2}$-type structure common
to most of the R$_{2}$T$_{2}$M (R=rare earths; T=transition metals;
M=Cd, In, Sn, and Pb). Comparison of the crystal structure with that
of other compounds reveals that the radii of the T and M atoms
together control the lattice parameters, and hence the overall
valence of the Yb ions, which are trivalent in Yb$_{2}$Pt$_{2}$Pb.
The crystal structure implies a layered magnetic structure, with the
Yb ions contained in two different types of planes. In each, the
moments lie on contiguous isosceles triangles, which can be mapped
onto the Shastry-Sutherland model. Strong magnetic anisotropy is
found at low temperatures with
$\chi_{\lbrack100\rbrack}/\chi_{\lbrack001\rbrack}\sim30$,
indicating that the magnetic moments of the Yb ions are confined to
planes, and likely lie along the $\lbrack110\rbrack$ and
$\lbrack1\bar{1}0\rbrack$ directions. The magnetic structures
implied by the magnetization measurements feature planes of
orthogonal Yb dimers which lead to spin liquid ground states in
other Shastry-Sutherland lattice systems.

Long range antiferromagnetic order occurs at 2.07 K, emerging from a
paramagnetic state which magnetic susceptibility, electrical
resistivity, and specific heat measurements find to have substantial
short range order and frustration. These experimental findings
suggest that Yb$_{2}$Pt$_{2}$Pb may be a rare example of a
Shastry-Sutherland lattice system, although additional experimental
work is required to demonstrate this definitively.

\begin{acknowledgments}
The authors are grateful to C. Broholm and P. Schiffer for useful
discussions and to C. Henderson for assistance with the
electron-probe microanalysis, which was performed at the University
of Michigan Electron Microbeam Analysis Laboratory (EMAL). Work at
the University of Michigan and at Stony Brook University is
supported by the National Science Foundation under grant
NSF-DMR-0405961.

\end{acknowledgments}

\end{document}